\documentclass[12pt, letterpaper]{article}

\usepackage{amsmath, amssymb, graphics, epsfig, graphicx, color}
\usepackage{epsf}
\usepackage{epstopdf}

\newcommand{\nc}{\newcommand}
\nc{\ba}{\begin{eqnarray}}
\nc{\ea}{\end{eqnarray}}

\newcommand{\calR}{{\cal{R}}}

\newcommand{\calP}{{\cal{P}}}

\def\bfk{{\bf k}}

\begin{document}

\vspace{5mm}
\vspace{0.5cm}
\begin{center}

{\Large \bf Effective Field Theory of non-Attractor Inflation}
\\[0.5cm]

{\large Mohammad Akhshik$^{1, 2}$, Hassan Firouzjahi$^2$,  Sadra Jazayeri$^{1}$}

{\small \textit{$^1$Department of Physics, Sharif University of Technology, Tehran, Iran
}}\\
{\small \textit{$^2$School of Astronomy, Institute for Research in Fundamental Sciences (IPM) \\ P.~O.~Box 19395-5531, Tehran, Iran
}}

\end{center}

\vspace{.8cm}


\begin{abstract}

We present the model-independent studies of  non attractor inflation in the context of effective field theory (EFT) of inflation.  
Within the EFT approach two independent branches of non-attractor inflation solutions are discovered in which  a near scale-invariant curvature perturbation power spectrum is generated from the interplay between the variation of sound speed and the second slow roll parameter $\eta$.  The first branch 
captures and  extends the previously studied models of non-attractor inflation in which the curvature perturbation is not frozen on super-horizon scales and the single field non-Gaussianity consistency condition is violated. We present the general expression for the amplitude of local-type non-Gaussianity in this branch.  The second branch is new in which  the curvature perturbation is frozen on super-horizon scales and the single field non-Gaussianity consistency condition does hold in the squeezed limit. Depending  on the model parameters,  the  shape of bispectrum in this branch 
changes from  an equilateral configuration to a folded configuration while  the amplitude of 
non-Gaussianity  is less than unity.

\end{abstract}

\newpage
\section{Introduction}

The basic predictions of models of inflation are well consistent with cosmological observations 
\cite{Ade:2013lta, Ade:2013uln}. The large scale curvature perturbations generated from quantum fluctuations during inflation are nearly scale-invariant, nearly adiabatic and nearly Gaussian as verified
very accurately in recent cosmological observations.  
Despite the immense success of inflation as the leading paradigm for early universe and structure formation there is no unique theory of inflation.  There are many models of inflation consistent with data. The simplest models of inflation are based on the dynamics of a
scalar field rolling slowly over a near flat potential. Even in the context of single field inflationary models, there are numerous scenarios of inflation. Therefore, it is an important question how far one can capture the  most robust predictions of models of inflation without relying on particular realization of inflation. 

Effective Field Theory (EFT) of inflation \cite{Cheung:2007st} has provided a good answer to this question. In EFT language, all interactions compatible with the underlying symmetries are 
allowed. Then  different models of inflation are realized depending on how one turn on particular interactions in the effective action governing the dynamics of  the light field. So far the best studied set up of  EFT of inflation is the single field models of inflation in which only one field is responsible for generating the curvature perturbations. However, the picture of EFT of inflation can be extended to more complicated setups such as multiple field scenarios \cite{Senatore:2010wk}. 

One interesting applications of EFT of inflation is its ability to classify different models of inflation based on their predictions for non-Gaussianity. Although there is no detection of non-Gaussianity from
the Planck data \cite{Ade:2013ydc}, but non-Gaussianity continues to play important roles in constraining models of inflation at an age of precision  cosmology  \cite{Alishahiha:2004eh, Komatsu:2009kd, Chen:2010xka, Komatsu:2010hc}.  In particular, it is known that models of single field slow-roll inflation follow the Maldacena's  consistency condition \cite{Maldacena:2002vr, Creminelli:2004yq} in which the amplitude of non-Gaussianity $f_{NL}$ in the squeezed limit is related to the spectral index of curvature perturbation power spectrum $n_s$ via $f_{NL}^{ \mathrm{sq} }  \sim n_s -1$.\footnote{ It is worth mentioning that, as argued in \cite{Pajer:2013ana}, the freely falling observers in Fermi normal coordinates can not observe this small amount of non-Gaussianity
as bispectrum in these sets of coordinates is cancelled up to corrections of ${O}(k_L^2/k_S^2)$ in which $k_L$ and $k_S$ are the long mode and the short mode defined in squeezed configuration.} This has led to the conclusion that a detection of local type non-Gaussianity at the order $f_{NL} = O(1)$ will rule  out {\it all  } single field models of inflation. However, it was shown in \cite{Namjoo:2012aa} that this general statement does not hold in models of non-attractor inflation in which the slow-roll parameter  $\epsilon \equiv -\frac{\dot H}{H^2}$, measuring the variation of the Hubble parameter $H$,   falls off exponentially during inflation. As a result, the would-be decaying mode of the curvature perturbations actually grows during the non-attractor phase and the curvature perturbations $\calR$
is not frozen on super-horizon scales.  Therefore, the single field non-Gaussianity
consistency condition is violated 
and $f_{NL} \geq O(1)$ can be generated during the non-attractor phase \cite{Namjoo:2012aa, Chen:2013aj, Chen:2013eea}.  

So far there are few known examples of non-attractor scenarios. The simplest model is presented in \cite{Namjoo:2012aa}, see also \cite{Kinney:2005vj, Motohashi:2014ppa}, in which 
the scalar field is rolling in a constant potential with sound speed $c_s=1$. A different model of non-attractor inflation was presented in \cite{Chen:2013aj, Chen:2013eea} in the context of $P(X, \phi)$ setup of K-inflation  \cite{ArmendarizPicon:1999rj, Garriga:1999vw},  see 
also \cite{Huang:2013oya},  with a constant sound speed $c_s$. With the motivation of EFT of inflation, it is a natural question to ask what are the generic predictions of
 non-attractor models of inflation without relying on a particular realization of non-attractor scenario?
 The goal of  this paper is to address this question. As we shall see, the non-attractor scenarios can easily be embedded within the general context of EFT of inflation. Equipped with the power of EFT of inflation, we present the general predictions of the non-attractor scenarios for the spectral index and the shapes of non-Gaussianities. We demonstrate that there are two independent branches of non-attractor models which can generate a nearly scale-invariant power spectrum. In one branch we capture and extend the previous results obtained in \cite{Namjoo:2012aa, Chen:2013aj, Chen:2013eea}. Very 
 interestingly,  the other branch of  non-attractor scenarios is new which escaped the previous model-dependent studies.

\section{The non-attractor inflation in EFT setup}

In this section we present our setup of non-attractor inflation in the context of EFT of inflation.
Let us first review the EFT of inflation very briefly. An extensive discussion can be found in 
\cite{Cheung:2007st}.  In what follows, we use the conventions and the methods employed in 
\cite{Cheung:2007sv}. 

In a quasi-de-Sitter background with a time-dependent scalar field $\phi(t)$, the four-dimensional diffeomorphsim invariance  is spontaneously broken to a three-dimensional spatial diffeomorphsim invariance determined by  $\phi(t)=\mathrm{constant}$. In the so-called unitary (comoving) gauge, one is allowed to write down all terms in the action which is consistent with the three-dimensional diffeomorphsim invariance while yielding the known background for the Hubble expansion rate $H(t)$ and its derivative $\dot H(t)$. Alternatively, one can change this picture by restoring the explicit full four-dimensional diffeomorphsim invariance  by introducing a scalar field fluctuations, $\pi(x^{\mu})$, the Goldstone boson associated with the breaking of  the time diffeomorphsim invariance. 

In the presence of the perturbations, the metric is given by
\ba
\label{metric}
ds^2 = - N^2 d t^2 + \hat g_{ij} ( dx^i + N^i d t) ( dx^j + N^j dt)
\ea
in which $N(x^\mu)$ and $N^i(x^\mu)$ respectively represent the lapse and the shift functions and 
$ {\hat g}_{ij}$ represents the spatial metric. 

The full action, after restoring the four-dimensional diffeomorphsim invariance by the introduction of the Goldstone boson $\pi(x^\mu)$,  is $S_{\rm total }= S_{\rm matter}+ S_{\rm EH}$ in which the matter part of the action is   \cite{Cheung:2007sv}
\ba
\label{S-matter}
S_{\rm matter} = \int \! d^4 x  \sqrt{- g} &&\left[ -M^2_{P} \dot{H}(t+\pi)\left(\frac{1}{N^2}
\left(1+\dot\pi-N^i\partial_i\pi\right)^2-\hat
g^{ij}\partial_i\pi\partial_j\pi\right) \right. \nonumber\\
&&\left.
- M^2_{ P} \left(3H^2(t+\pi) +\dot{H}(t+\pi)\right)+ \right.\\
&&\left.\frac{M(t+\pi)^4}{2}\left(\frac{1}{N^2}
\left(1+\dot\pi-N^i\partial_i\pi\right)^2-\hat
g^{ij}\partial_i\pi\partial_j\pi-1\right)^2 + \right. \nonumber\\
\nonumber &&\left. \frac{c_3(t+\pi)\; M(t+\pi)^4}{6}
\left(\frac{1}{N^2} \left(1+\dot\pi-N^i\partial_i\pi\right)^2-\hat
g^{ij}\partial_i\pi\partial_j\pi-1\right)^3+ ... \right] \, .
\ea
In this picture the first two lines of Eq. (\ref{S-matter}) are fixed by the given form of the FRW background  while the term containing $M(t + \pi)$ in the third line  controls  the sound speed of scalar perturbations, $c_s$  \cite{Cheung:2007st}. More specifically, the relation between $M(t)$ and $c_s$ is given by 
\ba
\label{M-cs}
M^4=\frac{1}{2}\dot{H}M_p^2 \left(1-\frac{1}{c_s^2}\right) \, .
\ea
Note that both operators in the third and fourth lines of Eq. (\ref{S-matter})
containing $M(t + \pi)$ originates from the term $g^{00}+1$ in the unitary gauge (comoving gauge) and  $c_3(t+ \pi)$ is an undetermined coupling. As understood from the logics of EFT, there are many terms which are not written in the above action as they are not 
important in our limit of interest, the decoupling limit, in which the gravitational back-reactions are discarded and when we are well below the strong coupling limit. 

In addition, the gravitational part of the action from the Einstein Hilbert term is 
\ba
\label{EH} 
S_{\rm EH} = \frac 12 M_{P}^2 \int  d^4 x \,
\sqrt{-g} \, R = \frac 12 M_{ P}^2 \int \! d^3 x \, dt \:
\sqrt{\hat g} \, \big[ N R^{(3)} + \frac{1}{N} (E^{ij} E_{ij} -
E^i{}_i {}^2) \big] \, ,
\ea  
in which $M_P$ is the reduced Planck mass, $E_{ij}$ is related to the extrinsic curvature $K_{ij}$ associated with the three-dimensional hypersurface $\phi(t)={\rm constant}$ via $E_{ij} = N K_{ij}$  
and $R^{(3)}$ represents the three-dimensional Ricci scalar associated with the metric $\hat g_{ij}$. 

As discussed in  \cite{Cheung:2007st} the great advantage of EFT of inflation is in the decoupling limit in which one can safely neglect the gravitational back-reactions so the observational predictions such as the tilt of scalar perturbation  or the non-Gaussianity parameter $f_{NL}$ are determined to leading order  by the matter fluctuations. Technically, this corresponds to setting $N=1$ and $N^i=0$ in the action  and neglecting the  perturbations from the  gravitational action Eq. (\ref{EH}) compared to perturbations from  the matter action. Having this said,  in the appendix we integrate out the non-dynamical variables $N$ and $N^i$ and demonstrate the validity of the decoupling limit for our analysis.  

After presenting the setup of EFT, here we present our  definition of non-attractor scenarios. For this purpose, it is convenient to define the ``slow-roll'' parameters associated with the 
variations of $H(t)$  via
\ba
\label{epsilon-eta-def}
\epsilon \equiv -\frac{\dot H}{H^2} \quad , \quad \eta \equiv \frac{\dot \epsilon}{\epsilon H}  \, .
\ea
Our general definition of the non-attractor scenarios is that during non-attractor phase the slow-roll 
parameter $\epsilon$ falls off exponentially such that the background becomes more and more like a dS background as inflation proceeds. Consequently, the second slow-roll parameter $\eta$ is not small.
Indeed, we expect $\eta \sim O(-1)$ during the non-attractor phase. Not that the requirement that $\epsilon$ falls off implies that $\eta <0$. The exact value of $\eta$ is determined from $n_s$. For example, in non-attractor models studied in \cite{Namjoo:2012aa, Chen:2013aj, Chen:2013eea} in which the sound speed is constant, $n_s$ is given by $n_s -1 = 6+\eta$ so requiring a near scale-invariant power spectrum  to be consistent with  observations implies $\eta \simeq -6$.
In these models, with $\eta \simeq -6$, we find that $\epsilon$ falls off like $1/a(t)^6$. As we shall see, during this non-attractor phase the curvature perturbation $\calR$ is not frozen on super-horizon scales
and it grows like $a(t)^3$. This is the main reason why the celebrated  Maldacena's consistency condition is violated in non-attractor model.  We stress that  the non-attractor phase can not extend for a long period, as $\epsilon$ becomes very small while $\calR$ grows exponentially and  the system becomes non-perturbative. In order to prevent this to happen, the non-attractor phase has to be followed by an attractor phase in which $\epsilon$ becomes nearly time-independent and $\calR$ saturates on super-horizon scales. 

Motivated with the generality of EFT, here we extend the non-attractor models to case in which the sound speed and the operator $c_3(t)$ can also vary during the non-attractor phase. For this purpose, let us define the slow-roll parameters in the matter sector via
\ba
\label{dot-c}
s\equiv \frac{\dot c_s}{H c_s} \quad , \quad \delta_3 \equiv \frac{\dot c_3}{H  c_3} \, .
\ea  
In principle the parameters $s$ or $\delta_3$ can take arbitrary values as long the 
power spectrum becomes nearly scale-invariant. 

With these discussions one may ask what conditions should be imposed on the variations of the secondary slow-roll parameters $\eta, s$ and $\delta_3$. We take the simple assumption that these secondary slow-roll parameters are nearly constant and $\dot \eta \sim \dot s \sim  \dot \delta_3 \sim O(\epsilon)$. Of course, this may not be the case in general and these secondary slow-roll parameters may have rapid variations like $\epsilon , c_s$ or $c_3$,  but as we shall see, even the simple assumptions of allowing  $|\eta|, |s| , |\delta_3| \sim O(1) $ while neglecting their evolution describes large enough class of non-attractor models.  

This discussion summarizes our definition of non-attractor phase without relying on any particular model. Our goal is to understand the properties of non-attractor models in a model-independent way using the effective field theory (EFT) method of inflation \cite{Cheung:2007st}. As we shall see our model-independent analysis reproduces the results obtained in \cite{Namjoo:2012aa, Chen:2013aj, Chen:2013eea} in particular limits. 

Now the question arises how the non-attractor models can be embedded in EFT approach of inflation.
Interestingly, this is vary simple. During the non-attractor phase $\epsilon$ falls off rapidly so the 
the main places where the effects of non-attractor phase enter are the first two lines of action (\ref{S-matter}) containing $\dot{H}(t+\pi)$ and $H^2(t+\pi)$. Therefore, we need to expand 
$\dot{H}(t+\pi)$ and $H^2(t+\pi)$ to  first order of $\epsilon$. Note that the presence of 
$\eta \simeq O(-1)$ will play a crucial role here. We have
\ba
\ddot{H}&=&-\dot{\epsilon}H^2-2\epsilon \dot{H}H \nonumber \\
&=&-\eta \epsilon H^3 + 2\epsilon ^2 H^3
= -\eta \epsilon H^3 + O(\epsilon ^2),
\ea
and
\ba
\frac{d \ddot{H}}{dt}=-\dot{\eta}\epsilon H^3 - \eta ^2 \epsilon H^4+7\eta \epsilon ^2 H^4-6\epsilon ^3 H^4 = -\eta ^2 \epsilon H^4 + O(\epsilon ^2)
\ea
Note that in the last relation we have neglected the evolution of $\eta$ because 
$ \frac{\dot{\eta}}{\eta H} = O(\epsilon) $ as assumed above.

Also,
\ba
\frac{d^2}{dt^2}\ddot{H}=-\eta ^3 \epsilon H^5 + O(\epsilon ^2).
\ea

Combining these expansions for $\frac{d \ddot{H}}{dt}$ and $\frac{d^2}{dt^2}\ddot{H}$ to leading order for $H^2(t+\pi)$ we obtain 
\begin{align}
\label{H2}
H^2(t+\pi)&=H^2(t)+2H\dot{H}\pi +\frac{1}{2}\left(\dot{H}^2+2H\ddot{H}\right)\pi ^2+\frac{1}{6}\left(6\dot{H}\ddot{H}+2H\frac{d\ddot{H}}{dt}\right)\pi ^3+ \ldots  \nonumber \\
&=H^2(t)-2\epsilon H^3 \pi -\eta \epsilon H^4 \pi ^2 - \frac{1}{3}\eta ^2 \epsilon H^5 \pi ^3 +\ldots \, . 
\end{align}
Similarly,  for $\dot{H}(t+\pi)$ to leading order we obtain
\ba
\label{H-dot}
\dot{H}(t+\pi)=-\epsilon H^2-\eta \epsilon H^3 \pi -\frac{1}{2}\epsilon \eta ^2 H^4 \pi ^2 -\frac{1}{6}\eta ^3 \epsilon H^5 \pi ^3 + \ldots \, .
\ea
We stress that in obtaining the above expansions, we have kept terms leading in $\epsilon$ because 
$\epsilon$  falls off rapidly during the non-attractor phase. In addition we have 
neglected the variations of $\eta$ as we assumed that $\dot \eta \sim O(\epsilon)$.  

In addition to contributions from $H^2(t + \pi)$ and $\dot H(t+ \pi)$ from the background dynamics, we also have the contributions from the rapid variations of the sound speed $c_s(t+ \pi)$ and the 
coupling $c_3(t+ \pi)$. These contributions should also be included in quadratic and cubic actions. 
As we discussed before, the operator $M(t)$ controls the sound speed of scalar fluctuations 
as given in Eq. (\ref{M-cs}).  As a result
\ba
\label{M4}
M^4(t + \pi) = -\frac{1}{2}\epsilon H^2 M_P^2  \left[ \left(1-\frac{1}{c_s^2}\right)  + \frac{2 s}{c_s^2}  H \pi  + \eta   \left(1-\frac{1}{c_s^2}\right) H  \pi  \right]  + ...
\ea
Similarly, for the operator $c_3(t)$ we have 
\ba
\label{c3}
c_3(t + \pi) = c_3(t) \left(  1+ \delta_3 H \pi \right) + ... 
\ea
Equipped with the expansion of the background parameters $H(t+ \pi)$ and $\dot H(t+ \pi)$
as given in Eqs. (\ref{H2}) and (\ref{H-dot}) and the matter sector parameters 
Eqs. (\ref{M4}) and (\ref{c3}) we can obtain the quadratic and cubic action for $\pi$ in the matter sector along with the action from the gravitational sector from Eq. (\ref{EH}). 

As we mentioned before, the advantage of EFT of inflation is in decoupling limit in which one can neglects the gravitational back-reactions. This is even more justified in the non-attractor regime in which $\epsilon$ falls off rapidly and the errors in neglecting the gravitational back-reactions which are controlled by higher powers of $\epsilon$ are quite negligible. As we see from the detail analysis in the Appendix, we can safely go to decoupling limit in which we can 
ignore the contribution from  Eq. (\ref{EH})
and set $N=1, N^i=0$ in the matter sector. In this limit, the quadratic and the cubic actions
generated from Eq. (\ref{S-matter})  respectively   are 
\begin{equation}
\label{action-quad}
S_2=\int d^3x\, dt\, M_P^2a^3\epsilon H^2\left(c_s^{-2}\dot{\pi}^2-a^{-2}(\partial _i \pi)^2\right) \, ,
\end{equation}
and
\ba
\label{S3}
S_3 &=& \int d^4x a^3 \epsilon M_p^2 H^2 \Big[ \frac{(\eta-2s)H}{c_s^2}\pi \dot{\pi}^2 - \frac{\eta H}{a^2}\pi (\partial_i \pi)^2 \\ \nonumber
&&~~~~~~~~~~~~~~~~~~~~~~~~-(1+\frac{2}{3}c_3)(1-\frac{1}{c_s^2})\dot{\pi}^3+(1-\frac{1}{c_s^2}) \frac{\dot{\pi} }{a^2} (\partial_i \pi)^2 \Big]  \, .
\ea

The above actions are written in terms of $\pi$. However, we are interested in power spectrum and bispectrum of comoving curvature perturbation  $\calR$. The relation between $\pi$ and $\calR$, to quadratic order which is necessary for the bispectrum analysis, have been worked out in \cite{Maldacena:2002vr, Cheung:2007sv} yielding 
\ba
\label{R-pi}
\calR =  - H \pi + H \dot \pi \pi + \frac{\dot H}{2} \pi ^2 + O(\pi^3) \, .
 \ea 
We have specifically  checked that the above relation does hold for the non-attractor setups too.

Finally, to calculate the quadratic and cubic actions and the follow up in-in analysis we need to know the functional form of
the quantities $\epsilon(t), c_s(t) $ and $ c_3(t)$. Using the definition of $ \eta, s$ and $\delta_3$ given in Eqs. (\ref{epsilon-eta-def}) and  (\ref{dot-c}), and employing the assumption that $\eta, s$ and $\delta_3$ do not vary during the non-attractor phase  we have 
\ba
\label{epsilon-c-back}
\epsilon = \epsilon _e \left( \frac{\tau}{\tau _e}\right)^{-\eta} \quad , \quad 
 c_s=c_{s  e}\left(\frac{\tau}{\tau _e}\right)^{-s}  \quad , \quad 
 c_3 = c_{3  e} \left( \frac{\tau}{\tau _e}\right)^{-\delta_3} \, ,
\ea 
in which $\tau$ is the conformal time related to cosmic time via $d\tau = dt/a(t)$ and  $\tau _e$  represents the time of end of 
non-attractor phase.  Finally, the scale factor in terms of conformal time is given by $a= a_e \left(\frac{\tau}{\tau _e}\right)^{-1}$.

As we shall see, a combination of $\eta$ and $s$ determine the tilt of curvature perturbation power spectrum. Therefore, neither $\eta$
nor $s$ is fixed individually. Having this said, we can derive an upper bound on the value of $s$ as follows. The universal definition of inflation is that $\ddot a >0$ so during this phase the comoving Hubble radius $1/aH$ falls off rapidly. However, in the inflationary model with a 
non-trivial sound speed, we encounter the ``sound horizon'' associated with the scalar perturbations. Correspondingly, the comoving sound horizon is given by $c_s/ a H$. In order for the perturbation with the comoving wave number $k$ to be sub-horizon during early stage of inflation and then to leave the comoving sound horizon during the subsequent stage of inflation,  we require that the comoving sound horizon  to be a decaying function during inflation. Using Eq. (\ref{epsilon-c-back}), the condition for the comoving sound horizon  $c_s/ a H$ to be a decaying function
as inflation proceeds is $s<1$ which will be imposed in our analysis below.

\section{Power spectrum}

In this section we study the predictions for curvature perturbations 
power spectrum. To leading order $\calR = -H\pi$ and the quadratic action for
curvature perturbations power spectrum is 
\ba
\label{S-2}
S_2=\frac{1}{2}\int d^3xd\tau z^2\left[ {\calR ^\prime}^2-c_s^2(\partial _i \calR)^2\right]  \, ,
\ea
in which a prime indicates the derivative with respect to conformal time, $\partial_i \calR$ represents the spatial derivative of $\calR$ 
and  the parameter $z$ is defined via
\ba
z^2=\frac{2\epsilon a^2}{c_s^2}M_P^2.
\ea
Interestingly the action (\ref{S-2}) coincides with action obtained in particular $P(X,\phi)$ model  of non-attractor inflation studied in  \cite{Chen:2013aj, Chen:2013eea}. In the decoupling limit 
in which we neglect the gravitational action (\ref{EH}) and, as long as we neglect the variation of $\eta$, the second order action for all non-attractor scenarios is uniquely given by Eq. (\ref{S-2}). The only relevant parameter  is the sound speed of perturbations $c_s(t)$, which is determined by the operator $M^4(t)$ given in Eq. (\ref{M4}). 

To quantize the system and to calculate the power spectrum, it is conventional to define the canonically normalized field $v$ related to
$\calR$ via  $ v = z \, \calR$. After some integrating by parts, the action for the canonically normalized field $v$ is given by 
\ba
S_2=\frac{1}{2}\int d^3xd\tau \left[\, (v^\prime)^2-c_s^2(\partial _i v)^2+\frac{z^{\prime\prime}}{z}v^2 \, \right],
\ea
in which the quantity $\frac{-z''}{z}$ represents the effective time-dependent mass of the canonically normalized field. More specifically,  to leading order of $\epsilon$ we have 
\ba
\frac{z^{\prime\prime}}{z}=\frac{1}{\tau ^2}\left( 2+\frac{1}{4}(\eta - 2s)(\eta -2s +6)\right) \, .
\ea

The equation of motion for the canonically normalized field  in the Fourier space with the momentum number $\bfk$ is 
\ba
\label{v-eq}
v_k'' + \left( c_s^2 k^2 - \frac{ \nu^2 -\frac{1}{4} }{\tau^2} \right) v_k =0 \, ,
\ea
in which a prime indicates the derivative with respect to the conformal time and  
the index $\nu$ is defined via 
\ba
\label{nu-def}
\nu \equiv \frac{\eta - 2 s + 3}{2 } \, .
\ea
We present the exact solution of $v_k(\tau)$ in Eq. (\ref{v-eq}) for the form of $c_s$ given in Eq. (\ref{epsilon-c-back}). However, before presenting the general solution for $v_k(\tau)$ let us pause and check whether or not  the notion of propagating wave and the initial adiabatic mode is well-defined in our system when $|s|$ is not small and $c_s(\tau)$ evolves rapidly. 
In order to have a well-defined propagating mode deep inside the horizon, we impose the WKB approximation in which $ | c_s' k/(c_s k)^2 | \ll 1 $. For this to apply we require 
\ba
\label{WKB-cond}
 c_s k |\tau| \gg |s| \, .
\ea
In this limit, one can neglect the second term in the big bracket  in Eq. (\ref{v-eq}) 
and a WKB solution for modes propagating deep inside the horizon is applicable. The corresponding 
WKB solution is
\ba
\label{WKB-sol}
v(k, \tau )  &\simeq& \frac{1}{\sqrt{2 k c_s}} \exp \left[ - i k \int c_s(k, \tau) d \tau
\right] \nonumber\\ 
&\simeq& \frac{1}{\sqrt{2 k c_s}} \exp \left( 
\frac{-i k\, c_{s}(\tau)  \tau}{1-s} \right )
 \qquad \qquad   \left(  c_s k |\tau| \gg |s|  \right)  \, .
\ea
Note that the only change compared to standard case in which one can neglect the evolution of $c_s$ is the appearance of the additional factor $1-s$ in the denominator of the argument.  As we discussed before, in order to have a decreasing function of comoving 
sound horizon, we require $1-s >0$.

Happily Eq. (\ref{v-eq}) can be solved analytically. The solution which matches the WKB solution 
for modes deep inside the horizon Eq. (\ref{WKB-sol}) is
\ba
v = C_1 \sqrt{x} H_{\mu}^{(1)} \left( \frac{x}{1-s} \left( \frac{x}{x_*} \right)^{-s}
\right)
\ea
in which  $H_{\mu}^{(1)}(x)$ is the Hankel function of type one, 
$x\equiv - c_s(\tau_e)  k \tau$ and 
\ba
\label{C1-def}
\mu \equiv  \frac{\nu}{1-s} \quad , \quad 
C_1 \equiv \sqrt{\frac{\pi}{4 k c_* ( 1- s) }}\,   e^{i(\mu + 1/2)\pi /2} \, .
\ea

We are interested in the power spectrum of the curvature perturbation $\calR = \frac{c_s}{\sqrt{2 \epsilon } a M_P} v$ on super-horizon scales in which $ x \rightarrow 0 $. Using the following formula for the small argument of Hankel function
\ba
H_{\mu}^{(1)} ( x) = e^{-i ( \mu - | \mu| ) \pi/2} H_{|\mu|}^{(1)} ( x)  \simeq 
e^{-i ( \mu - | \mu| ) \pi/2} \left[ \frac{- i \Gamma(|\mu|)}{\pi} \left( \frac{2}{x} \right)^{|\mu|} \right] 
\quad  \quad ( x \rightarrow 0 )
\ea
the curvature perturbation on super-horizon scales is obtained to be
\ba
\label{R-abs}
\Big|\calR (  \tau \rightarrow 0)  \Big | \simeq  \frac{ 2^{|\mu| - \frac{3}{2}} \Gamma(|\mu|)   (1- s)^{|\mu| - \frac{1}{2}}}{ \sqrt{ \pi \epsilon_e c_{se} k^3} } \frac{H}{M_P}  
\, \left( \frac{\tau}{\tau_e}
\right)^{ \nu + s |\mu|  -\frac{3}{2}} \left( - c_{s} k \tau \right)^{\frac{3}{2}- | \mu|}
\ea
in which $\epsilon_e$ and $c_{s e}$ represents the value of the corresponding quantities at the end of
non-attractor phase $\tau =\tau_e$ and $\Gamma$ is the Gamma function.
 
Correspondingly, the power spectrum at the end of non-attractor is obtained to be
\ba
\label{power}
\calP_\calR(\tau_e)  \equiv  \frac{k^3}{2 \pi^2}  \Big|\calR (  \tau_e)  \Big |^2 = 
{\cal A} \left( \frac{k}{k_e} \right)^{3 - 2 | \mu| } \, ,
\ea
in which $k_e \equiv a (\tau_e) H/c_{s e}$ represents the mode which leaves the sound horizon at the end of non-attractor phase and  the amplitude $ {\cal A}$  (COBE normalization) is
\ba
 {\cal A} \equiv \frac{\Gamma(|\mu|)^2 ( 1- s) ^{2 |\mu| -1} }{ \pi^3 2^{4 - 2 |\mu|}} \frac{1}{\epsilon_e c_{se}} \left(  \frac{H}{M_P} \right)^2 \, .
\ea

The spectral index $n_s$ from the power spectrum (\ref{power}) is obtained to be 
$n_s -1 = 3 -2 | \mu| $. Depending on the sign of $\mu$ we have 
\ba
\label{ns-eq1} 
n_s -1 =  -\frac{\eta+ s}{1-s}  \quad \quad ( \mu >0) 
\ea
or
\ba
\label{ns-eq2} 
n_s -1 =  \frac{6+ \eta- 5 s}{1-s}  \quad \quad ( \mu <0)  \, .
\ea
The branch represented by  Eq. (\ref{ns-eq1}) reproduces the results in the known slow-roll limit, the K-inflation model \cite{Garriga:1999vw,  Chen:2006nt, Shandera:2006ax},  in which $| s|, | \eta| \ll 1$ and  $n_s -1 \simeq - \eta - s - 2 \epsilon$ . Note that we have already discarded $\epsilon$ in our analysis assuming that $\epsilon \rightarrow 0$  so the $O (\epsilon)$
discrepancy with the K-inflation results is expected.  Furthermore, the branch represented by  Eq. (\ref{ns-eq2}) is a generalization of  the previously studied models of non-attractor with $s=0$ \cite{Namjoo:2012aa, Chen:2013aj, Chen:2013eea} 
in which $\eta \simeq -6$ to get a nearly scale-invariant power spectrum. Note that the branch $\eta \simeq -s$ represented  by Eq. (\ref{ns-eq1}) is  new which was not noticed in the previous model-based analysis of \cite{Namjoo:2012aa, Chen:2013aj, Chen:2013eea}. This demonstrates the power of  EFT as a platform to study inflation model independently. We comment that the condition
$\eta =-s$ to obtain a scale-invariant power spectrum may be interpreted from the fact that $\calP_\calR \propto 1/\epsilon c_s$. So, one may expect that a variation of $c_s(t)$ should be balanced by
the variation of $\epsilon(t)$ such that the power spectrum remains scale invariant when $\eta =-s$.

Equipped with the powerful EFT method  we see that  there are different options to get a nearly scale-invariant power spectrum with a mild red-tilted power, $n_s \simeq 0.96$, depending on whether $\mu >0$ or $\mu<0$.  Note that our starting assumption was that $\epsilon$ falls off rapidly so we choose the sign of $\eta$ to be negative. As for $s$, as we discussed before, we only require $s<1$ in order to get a decreasing function of comoving sound horizon during inflation so the sign of $s$ is  undetermined. A positive (negative) $s$ indicates a growing  (decaying) sound speed during inflation. In the case $s>0$ the sound speed can cross unity and we enter the superluminal regime. In order for this not to happen, one concludes that the duration of the non-attractor phase to be finite, say few e-folds such that  $ | s N_e| <1$ in which $N_e$ represents the duration of non-attractor phase. On the other hand, if $s<0$ then the sound speed falls off as inflation proceed. If $c_s$ becomes arbitrarily small, then one encounters the strong coupling regime limit 
with very large non-Gaussianity. Therefore, in order to avoid the strong coupling limit, we demand that the non-attractor phase terminates before $c_s$ becomes very small, say $c_s > 0.003$   \cite{Cheung:2007st}. However, there is stronger upper bound on $c_s$ from the Planck's constrains on equilateral-type  non-Gaussianity, requiring $c_s \ge 0.02$ \cite{Ade:2013ydc}. 

Having obtained the two possible branches to get a nearly scale-invariant power spectrum as give by Eqs. (\ref{ns-eq1}) and (\ref{ns-eq2}), it is instructive to look at the evolution of $\calR$ on super-horizon scales. From Eq. (\ref{R-abs}), and noting that $|\mu | \simeq \frac{3}{2}$ to get a
scale-invariant power spectrum, we obtain  
\ba
\label{dot-R}
\dot \calR =  - \frac{( \eta + s) }{2} H\,  \calR    \qquad \qquad  ( k \tau \rightarrow 0 ) \, .
\ea
Interestingly, for the branch $\eta =-s$, the curvature perturbation is frozen on super-horizon scales. This is similar to the conventional models of single field slow-roll inflation. In addition, as discussed in  \cite{Creminelli:2004yq, Cheung:2007sv}, the constancy of $\calR$ is the essential reason for the validity of the single field  non-Gaussianity consistency condition. Therefore, in this branch we expect that the single field  non-Gaussianity consistency condition to hold
and $f_{NL}^{\rm (sq)}= O(\epsilon)$. We verify this conclusion explicitly in next Section. On the other hand, for the branch $\eta = -6 + 5 s$, we obtain $\dot \calR = 3 (1- s ) H \calR$ on super-horizon scales. This is in agreement with the results obtained in the particular  $P(X, \phi)$ model studied in \cite{Namjoo:2012aa, Chen:2013aj, Chen:2013eea} in which $s=0$ and $\dot \calR = 3 H \calR$ on super-horizon scales. As a result, since $\calR$ is not frozen on super-horizon scales, we expect that the   non-Gaussianity consistency condition to be violated
in this branch as we shall see explicitly from our in-in analysis. Note that since $s<1$ we conclude that  $\dot \calR >0$ and the curvature perturbations grows exponentially during the non-attractor phase in this branch. This indicates an instability so we have to terminate the non-attractor phase followed by an attractor phase  \cite{Namjoo:2012aa}.

As mentioned above, the branch $\eta= -6 + 5 s$ is an extension of the previous non-attractor models
based on $P(X, \phi)$ action as studied in \cite{Namjoo:2012aa, Chen:2013aj, Chen:2013eea}. Therefore, it is an interesting exercise to realize a specific model for the branch  $\eta =-s$ such as in  $P(X, \phi)$ setup. In principle one could have discovered the branch $\eta=-s$ from examples built in $P(X, \phi)$ setup. As experienced in \cite{Chen:2013aj} this requires careful inverse engineering. Having this said, we find it  interesting that the new branch $\eta=-s$ emerges naturally within the general  context of EFT of inflation.

\section{Bispectrum}

Having studied the power spectrum and the criteria to obtain a nearly scale-invariant power spectrum, here we study the predictions of the general non-attractor models for non-Gaussianity. Here the discussions become more interesting. First, EFT provides a framework in which different inflationary models can be classified based on their predictions for the amplitude and the shape of non-Gaussianity. Second, the  particular non-attractor  models presented originally in \cite{Namjoo:2012aa, Chen:2013aj, Chen:2013eea} are the very few known examples of single field inflation which violate the   non-Gaussianity  consistency condition, for more discussions on non-Gaussianity  consistency condition see \cite{Maldacena:2002vr, Creminelli:2004yq, Cheung:2007sv,  
Assassi:2012et, Flauger:2013hra, Berezhiani:2014tda, Berezhiani:2013ewa, Sreenath:2014nca, Sreenath:2014nka, Creminelli:2012ed, Creminelli:2013cga, Dimastrogiovanni:2014ina}.  Therefore, it is natural to ask if this violation of
non-Gaussianity consistency condition is a generic feature of the non-attractor inflationary systems. 

To calculate the bispectrum of $\calR$ we have to take into account the non-linear relations between $\calR$ and $\pi$ given in Eq. (\ref{R-pi}). 
More specifically,  the bispectrum of $\calR$ is related to the bispectrum of
$\pi$ via
\ba
\label{pi-R-bi0}
\Big \langle \calR(\bfk_1)  \calR(\bfk_2) \calR(\bfk_2)  \Big \rangle =-H^3 \Big\langle \pi(\bfk_1)  \pi(\bfk_2) \pi(\bfk_3)  \Big\rangle
+ H^3 \Big \langle   \pi(\bfk_1) \pi(\bfk_2) \left(  \pi \dot \pi \right)({\bfk_3})   \Big\rangle  + 2 \mathrm{c.p.}
\ea
The last term above involves  convolution integrals, yielding  
\ba
\Big \langle   \pi(\bfk_1) \pi(\bfk_2) \left(  \pi \dot \pi \right)({\bfk_3})   \Big\rangle^{\prime}  =  \frac{1}{H^4} \left[ | \calR(\bfk_1) |^2 \calR({\bfk_2} )
{\dot \calR({\bfk_2}) }^* +   |\calR(\bfk_2) |^2 \calR({\bfk_1} ) {\dot \calR({\bfk_1}) }^*
\right] 
\ea
in which  $\langle \rangle^\prime$ indicates that we have absorbed the common factor 
$( 2 \pi)^3 \delta^{(3)} (\bfk_1 + \bfk_2 + \bfk_3) $. Now, using Eq. (\ref{dot-R}) to express $\dot \calR$ in terms of $\calR$, the contribution 
in bispectrum from the non-linear relation between $\pi$ and $\calR$ is
\ba
\label{pi-R-bi}
 H^3 \Big \langle   \pi_{\bfk_1} \pi_{\bfk_2} \left(  \pi \dot \pi \right)_{\bfk_3}   \Big\rangle^{\prime}  
 +2\mathrm{ c.p. }
=- (\eta + s) \Big[ P_{\calR}(\bfk_1)  P_{\calR}(\bfk_2)  +   P_{\calR}(\bfk_1)  P_{\calR}(\bfk_3) +  P_{\calR}(\bfk_2)  P_{\calR}(\bfk_3)
\Big] 
\ea
Plugging Eq. (\ref{pi-R-bi}) in  Eq. (\ref{pi-R-bi0}),  the relation between the bispectrum of $\pi$ and $\calR $ is
\ba
\label{pi-R-bi2}
\Big \langle \calR(\bfk_1)  \calR(\bfk_2) \calR(\bfk_2)  \Big \rangle^\prime  &=& \Big\langle \pi(\bfk_1)  \pi(\bfk_2) \pi(\bfk_3)  \Big\rangle^\prime
\nonumber\\ 
&-&  (\eta + s) \Big[ P_{\calR}(\bfk_1)  P_{\calR}(\bfk_2)  +   P_{\calR}(\bfk_1)  P_{\calR}(\bfk_3) +  P_{\calR}(\bfk_2)  P_{\calR}(\bfk_3)
\Big] 
\ea
in which it is understood that all quantities are calculated at the time of end of  non-attractor phase 
$\tau_e$.  The power spectrum  $P_{\calR}(\bfk)$  at  $\tau_e$ is
\ba
\label{powr-P}
P_{\calR}(\bfk, \tau_e) =  | \calR(\tau_e) |^2 = 
\frac{( 1- s)^2 H^2}{4 \epsilon_e c_e k^3 M_P^2} \, .
\ea

To calculate the first term above, $\Big\langle \pi(\bfk_1)  \pi(\bfk_2) \pi(\bfk_3)  \Big\rangle$, we need the cubic action which is given in Eq. (\ref{S3}). Specifically, using the standard in-in formalism \cite{Weinberg:2005vy, Chen:2010xka, Wang:2013zva}  the bispectrum  at the end of non-attractor phase is given by
\ba
\label{in-in}
\Big\langle \pi(\bfk_1, \tau_e)  \pi(\bfk_2, \tau_e) \pi(\bfk_2, \tau_e)  \Big\rangle  = i\int _{-\infty }^{\tau _e} d\tau  \Big\langle   \pi _{k_1}(\tau {_e})\pi _{k_2}(\tau{_e})\pi _{k_3}(\tau {_e}) {\cal L}_i(\tau) \Big \rangle + \mathrm{c.c.} 
\ea
in which c.c. stands for complex conjugation and 
${\cal L}_i $ represents either of the four Lagrangian terms in Eq. (\ref{S3});     namely
\ba
\label{L1}
\mathcal{L}_1&=& H^3 \frac{( \eta - 2 s) a^2}{c_s^2}\epsilon M_p^2 \pi { \pi ^\prime} ^2 \\
\label{L2}
\mathcal{L}_2 &=& -\eta H^3  a^2M_p^2\epsilon \pi (\partial _i \pi)^2, \\
\mathcal{L}_3&=&-{a}{H^2}\left(1+\frac{2}{3}c_3\right) \left(1-\frac{1}{c_s^2}\right) \epsilon M_p^2 {\pi ^\prime}^3,  \\
\mathcal{L}_4&=&{a}{H^2}\left(1-\frac{1}{c_s^2}\right)\epsilon M_p^2\pi ^\prime (\partial _i \pi)^2 \, .
\ea

Since the details of the analysis and the shape of non-Gaussianity are distinctly different for the two allowed branches of scenarios  $\eta = -6 + 5s$ and $\eta =-s$, we study each branch separately.


\subsection{ The branch $\eta =-s$}

In this case the wave function of $\calR$  is given by
\begin{equation}
\calR (k,\tau)=\frac{c_s}{2aM_p}\sqrt{-\frac{\pi \tau}{2\epsilon (1-s)}}H^{(1)}_{\frac{3}{2}}\left( -\frac{kc_s\tau}{1-s}\right) \, ,
\end{equation}
in which  we have neglected the corrections at the order $O(1-n_s)$ in wave function and set $\mu =\frac{3}{2}$. This is justified since we are interested in generating $f_{NL} \sim O(1)$ and can discard the  sub-leading corrections to $f_{NL}$ at the order $n_s-1$ or $\epsilon$. 

Happily the  in-in integrals can be performed  analytically. We present the result term by term. 
For the contribution from ${\cal L}_1$ we have 
\ba
\label{L1-bi}
\Big\langle \calR(\bfk_1)  \calR(\bfk_2) \calR(\bfk_3)  \Big\rangle^{\prime}_{{\cal L}_1}
 &=&\frac{ 3H^4 s(s-1)^4}{16M_P^4 K^2 k_1^3k_2^3k_3^3c_e^2\epsilon _e^2} 
 \\
 &\times&\Big[  \cos(2 \pi s) (k_1 k_2 k_3) ( k_1 k_2 + k_1 k_3 + k_2 k_3) + K (k_1^2 k_2^2 + k_1^2 k_3^2 + k_2^2 k_3^2 )  \Big], \nonumber
\ea
in which we have defined $K= k_1 + k_2 + k_3$. 

The contribution from ${\cal L}_2$ yields 
\ba
\label{L2-bi}
\Big\langle \calR(\bfk_1)  \calR(\bfk_2) \calR(\bfk_3)  \Big\rangle^{\prime}_{{\cal L}_2}  &= & \frac{H^4(k_1^2+k_2^2+k_3^2)}{32M_P^4 K^2} \frac{s(s-1)^4}{\epsilon _e ^2 c_e^2k_1^3k_2^3k_3^3}  \\
&\times& \Big[ \cos (2 \pi s) k_1 k_2 k_3  - K ( k_1^2 + k_2^2 + k_3^2 + k_1 k_2 + k_1 k_3 + k_2 k_3)  \Big] \, . \nonumber
\ea
Finally,  ${\cal L}_3$ and ${\cal L}_4$ make no contributions  $\langle \calR ^3 \rangle_{{\cal L}_3}=\langle \calR ^3 \rangle_{{\cal L}_4} =0$.

We also have to include the contributions from the non-linear relation between $\pi$ and $\calR$. 
Interestingly, in the current case in which $\eta+ s=0$, there is no contribution from the non-linear
relation between $\pi$ and $\calR$  as can be seen from Eq. (\ref{pi-R-bi}).
As a result, the total contribution to bispectrum is from ${\cal L}_1$ and ${\cal L}_2$.

To simplify the analysis, it is useful to define
\ba
\label{KPQ}
K\equiv k_1 + k_2 + k_3 \quad , \quad P^2 \equiv k_1 k_2 + k_1 k_3 + k_2 k_3 \quad , \quad
Q^3 \equiv k_1 k_2 k_3 \, .
\ea
With these definition, the total bispectrum is 
\ba
\label{bi-tot}
\langle \calR ^3 \rangle _{\mathrm{total} } ^\prime =\frac{s(s-1)^4 H^4}{32 M_P^4 \epsilon _e ^2 c_e^2 K^2 Q^9}   \Big[ Q^3 ( K^2 + 4 P^2) \cos( 2 \pi s) + K \Big( 4 P^4 - K^4 + 3 P^2 K^2 - 12 Q^3 K  \Big)
\Big] 
\ea
The above bispectrum has a non-trivial shape. However, from Eq. (\ref{dot-R}), 
we know that with $\eta + s= 0$, the curvature perturbation $\calR$ is frozen on super-horizon scales. Therefore, the  non-Gaussianity consistency condition should be satisfied in the squeezed limit \cite{Maldacena:2002vr, Creminelli:2004yq} in which $f_{NL}^{\mathrm{sq}} \propto (1- n_s) \rightarrow 0$. Therefore, it is constructive to look 
at the shape of bispectrum Eq. (\ref{bi-tot}) in the squeezed limit $k_3 \ll k_1 \simeq k_2$. In this 
limit $K\simeq 2 k_1, P^2 \simeq k_1^2 $ and  $Q\simeq k_3 k_1^2 \rightarrow 0$ and Eq. (\ref{bi-tot}) yields 
\ba
\label{bi-sq}
\langle \calR ^3 \rangle _{ \mathrm{total} } ^{ \prime \mathrm{sq} } \,  \simeq s \frac{k_3}{k_1} P_\calR (k_1) P_\calR(k_3) 
\rightarrow 0 \, ,
\ea
so the bispectrum vanishes in the squeezed limit, in agreement with the Maldacena's consistency condition. Note that we have neglected the gravitational back-reactions and also set $n_s=1$ in the
bispectrum analysis so we can not recover $O (\epsilon)$ corrections in bispectrum in the squeezed limit.  

It is also instructive to look at the magnitude of non-Gaussianity in the equilateral limit $k_1 = k_2 =k_3$.  In this limit we obtain
\ba
{\langle \calR ^3 \rangle  ^{\prime \mathrm {eq}} } = \frac{7}{6}s \cos (2 \pi s) P_\calR^2 (k_1) \, .
\ea
A convenient way to parameterize the amplitude of non-Gaussianity is via the parameter $f_{NL}$
defined via
\ba
\label{fNL-def}
\frac{ {6}}{5} f_{NL} = \frac{\Big\langle \calR(\bfk_1)  \calR(\bfk_2) \calR(\bfk_3)  \Big\rangle^{\prime}}{P_{\calR}(k_1)  P_{\calR}({k_2}) + 2 \mathrm{c.p.}} \, .
\ea
Using this definition, the non-Gaussianity parameter in the equilateral configuration  $f_{NL}^{\mathrm{eq} }$
is obtained to be 
\ba
f_{NL}^{  \mathrm {eq}} = \frac{35}{108} s \cos (2\pi s)  \, .
\ea
Knowing that $s<1$ in order to have a decaying sound horizon during inflation, the above equation 
indicates that the  amplitude of $f_{NL}^{\mathrm{eq} }$ is less than unity so it is consistent with observational constraints. 

Now let us look at the shape of bispectrum in general case. Following the convention of Chen  \cite{Chen:2010xka}, the dimensionless shape function $S(k_1, k_2, k_3)$ is defined by 
$$ \langle \calR ^3 \rangle  = \frac{( 2 \pi)^4 \calP_\calR^2}{( k_1 k_2 k_3)^2} S(k_1, k_2, k_3)$$ yielding
\ba
\label{shape}
S(k_1, k_2, k_3) = \frac{s}{8} \Big[ \Big ( 1+ \frac{4 P^2}{K^2} \Big) \cos (2 \pi s) + 
\Big( \frac{4 P^4}{K Q^3} - \frac{K^3}{Q^3} + \frac{3 K P^2}{Q^3} - 12  \Big)
\Big] \, .
\ea
Note the curious form of the numerical factors. As can be seen from our shape plots, the numerical factors in the second big bracket above are  such that the bispectrum does not peak in the local shape. We note that the term containing $\cos (2 \pi s) $ generates the equilateral shape while  the terms in the second bracket generate the folded shape. Therefore, depending on the value of the parameter $s$, the shape of non-Gaussianity changes from the equilateral shape to folded shape.
The equilateral shape dominates when $| \cos (2 \pi s)| $ is large, i.e. when $s \sim 0$ or $s \sim 1$,
while the folded shape dominates when $\cos ( 2 \pi s) $ becomes small when $s \sim \frac{1}{2}$. 
This is similar to the phenomena observed in model of quasi single field inflation \cite{Chen:2009we, Chen:2009zp}, see also \cite{Emami:2013lma, Pi:2012gf}, in which the shape function evolves from the equilateral shape to local shape depending on the mass of the semi-heavy iso-curvaton field. Finally, the amplitude of  non-Gaussianity is less than unity for both equilateral and folded shapes so the predictions of the model for non-Gaussianity  are well-consistent with observational bounds. We stress again that the shape does not peak for local-type configuration which is supported from the fact that the  non-Gaussianity consistency condition is satisfied in the squeezed limit as
discussed in Eq. (\ref{bi-sq}).

\begin{figure}[!t]
  \includegraphics[width=0.45\textwidth]{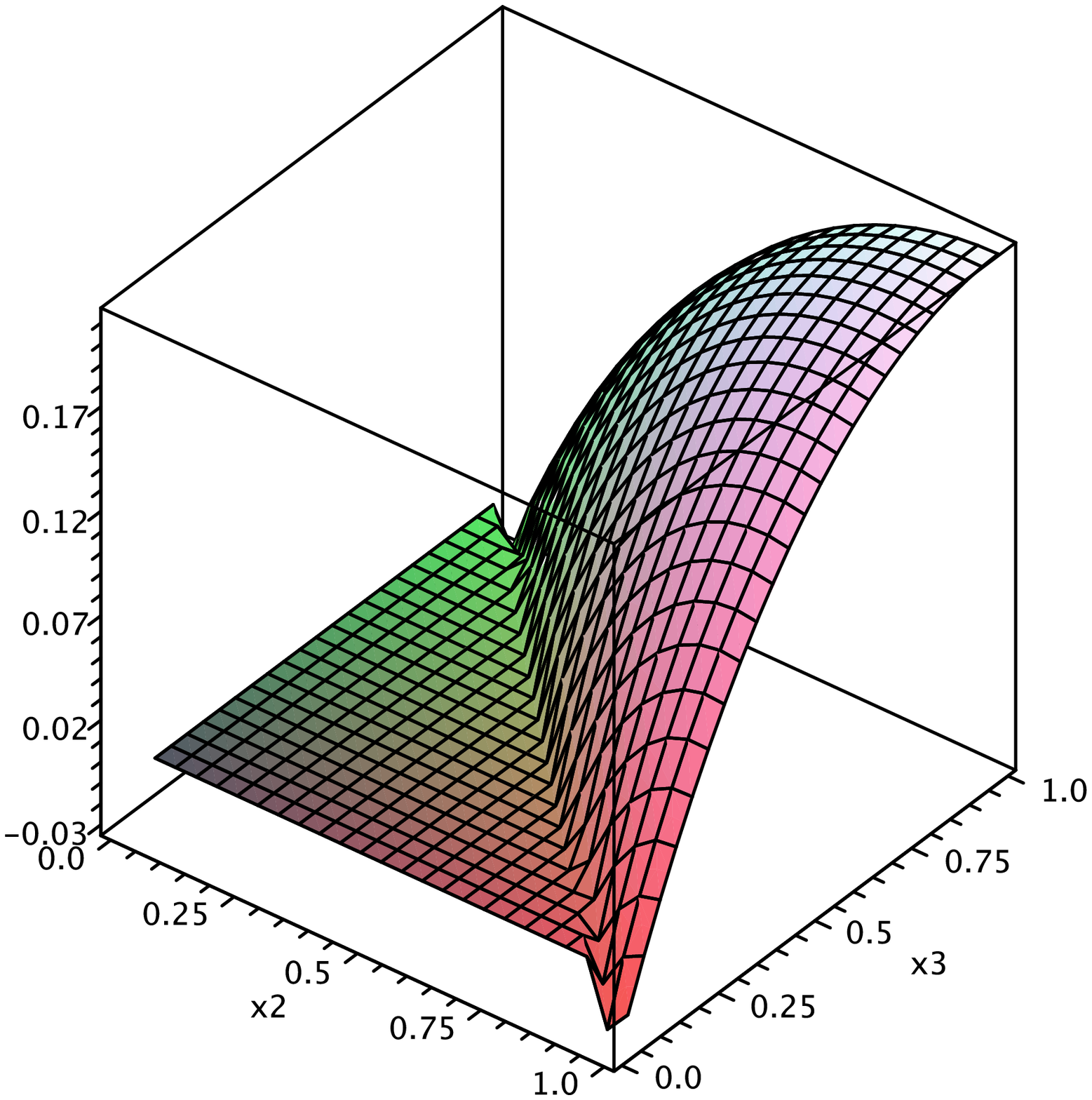}
  \hspace{0.03\textwidth}
  \includegraphics[width=0.49 \textwidth]{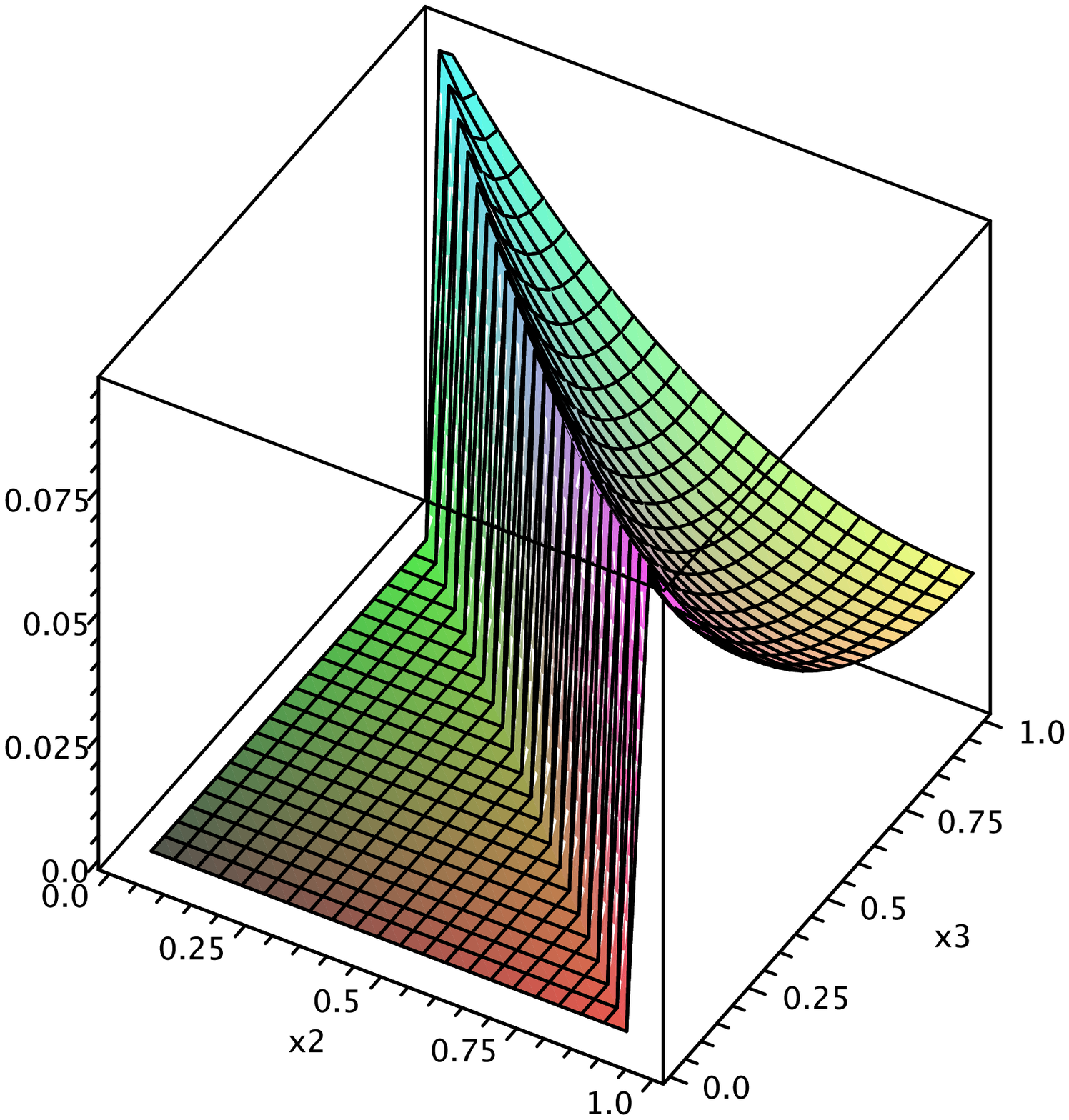}
  \caption{The amplitude of the shape functions, $|S |$,  given in Eq. (\ref{shape}). Left panel is for $s=0.9$ so the term containing $\cos (2 \pi s)$ dominates yielding a nearly equilateral shape. Right panel is for $s=0.3$ and the shape is close to the folded shape. Our convention is that $x_i= \frac{k_i}{k_1}$, for $i= 2, 3$. }
\end{figure}

\subsection{The branch $\eta =-6 + 5 s$}

Here we calculate the bispectrum for the case $\eta =-6 + 5 s$. Only ${\cal L}_1$ and ${\cal L}_3$ yield  non-zero contributions for in-in integrals and we have 
\ba
\langle \calR^3 \rangle^{\prime}_{ {\cal L}_1} = \frac{\eta - 2s}{32 } \frac{(s- 1 )^4H^4}{\epsilon_e^2 c_{s e}^2 M_P^4} \frac{k_1^3 + k_2^3 + k_3^3}{ k_1^3 k_2^3 k_3^3} 
\ea
and
\ba
\langle \calR^3 \rangle^{\prime}_{{\cal L}_3} = && \frac{(s- 1 )^6H^4}{\epsilon_e^2 c_{s e}^2 M_P^4}  \frac{k_1^3 + k_2^3 + k_3^3}{ k_1^3 k_2^3 k_3^3}  \nonumber\\
&\times& \Big[ \frac{9}{32}\,  \frac{6 ( 1- c_e^2) + 6 s c_{s e}^2 - 4 s}{(s-1) (4 s - 6)} + \frac{9 c_{3 e}}{8} \frac{(6- \delta_3) ( 1- c_{s e}^2) + 6 s c_{s e}^2 - 4 s }{ ( \delta_3 + 6 s - 6) ( 4 s + \delta_3 -6)}
\Big] \, ,
\ea
in which $c_{3e}$ represents the value of $c_3$ at the time $\tau =\tau_e$. 

The contribution from the non-linear relation between $\pi$ and $\calR$,  $\langle \calR^3 \rangle_{\pi \rightarrow \calR}$,   from Eq. (\ref{pi-R-bi})  is
\ba
\langle \calR^3 \rangle^{\prime}_{\pi \rightarrow \calR} = -( \eta + s) 
 \frac{(s- 1 )^4 H^4}{16 \epsilon_e^2 c_{s e}^2 M_P^4}  \frac{k_1^3 + k_2^3 + k_3^3}{ k_1^3 k_2^3 k_3^3} \, .
\ea
Adding the contributions from $\langle \calR^3 \rangle_{ {\cal L}_1}, \langle \calR^3 \rangle_{{\cal L}_3}$ and $\langle \calR^3 \rangle_{\pi \rightarrow \calR}$, the final bispectrum is obtained to be
\ba
\label{bi-total}
\langle \calR^3 \rangle^{\prime}_{total} =  \Big[ P(k_1) P(k_2) +  P(k_1) P(k_3) + P(k_2) P(k_3) \Big] {\cal F}
\ea
in which the function ${\cal F}$ is defined via
\ba
{\cal F} \equiv  \frac{ -( \eta + 4 s)}{2} + (s- 1)^2 \Big[    \frac{27 ( 1- c_{s e}^2) + 9s(3 c_{s e}^2 - 2 )}{(s-1) (4 s - 6)}  + 18 c_{3 e}   \frac{(6- \delta_3) ( 1- c_{s e}^2) + s( 6  c_{s e}^2 - 4 ) }{ ( \delta_3 + 6 s - 6) ( 4 s + \delta_3 -6)} \Big]  \nonumber
\ea
From Eq. (\ref{bi-total}) we see that the bispectrum has the exact local shape and no other shapes
are generated.  

Now let us calculate the non-Gaussianity parameter $f_{NL}$ in the squeezed limit $k_3 \ll k_1 \simeq k_2$. Using the general definition given in Eq. (\ref{fNL-def}), $f_{NL}$ in the 
squeezed limit is obtained to be $\frac{12}{5} f_{NL}^{\mathrm{sq}}= 2 {\cal F}$ yielding 
\ba
\frac{12}{5}  f_{NL}^{\mathrm{sq}} &=&  - ( \eta + 4 s) \\
&&+  (s- 1)^2 \Big[    \frac{54 ( 1- c_{s e}^2) + 18s(3 c_{s e}^2 - 2 )}{(s-1) (4 s - 6)}  + 36 c_{3 e}   \frac{(6- \delta_3) ( 1- c_{s e}^2) + s( 6  c_{s e}^2 - 4 ) }{ ( \delta_3 + 6 s - 6) ( 4 s + \delta_3 -6)} \Big]  \nonumber
\ea

As an example consider the $P(X, \phi)$ model presented in \cite{Chen:2013aj} and \cite{Chen:2013eea} with  $s= \delta_3 =0$ and $\eta =-6$. In this case we obtain
\ba
{\cal F} = 6 + 3 (1 -c_{s }^2) ( 2 c_3 + 3 ) \, .
\ea
Furthermore, if we assume $c_3 = -\frac{3}{2} + \frac{1}{2 c_s^2}$ which is valid in $P(X, \phi)$
model studied in  \cite{Chen:2013aj} and \cite{Chen:2013eea}
then ${\cal F} = \frac{3 ( 1 + c_s^2)}{c_s^2}$ and 
\ba
f_{NL} = \frac{5 ( 1+ c_s^2)}{ 4 c_s^2}  \, ,
\ea
in exact agreement with the results of \cite{Chen:2013aj} and \cite{Chen:2013eea}. Finally, if 
we further assume $c_s =1$, as in model studied in \cite{Namjoo:2012aa}, then $f_{NL}= \frac{5}{2}$ which is well-consistent with the Planck constraints on the amplitude of local-type non-Gaussianity \cite{Ade:2013ydc}.  

\section{Summary and Discussions}

In this paper we have presented a model-independent study of  non-attractor inflation in the context of EFT of inflation. The goal was to find the generic predictions of the non-attractor scenarios for the power spectrum and bispectrum. We believe  that it is important to understand the general predictions of the non-attractor  scenarios in a model-independent way. This is mainly because the non-attractor scenarios provide the very few examples in which the Maldacena's non-Gaussianity 
consistency condition for single field inflation is violated. As a result, a detection of local-type non-Gaussianity with $f_{NL} \sim O(1)$ does not necessarily rule out {\it all} single field models of inflation. More precisely, it rules out all single field models of inflation which has reached the attractor phase.

We have studied a large class of non-attractor model in the context of EFT of inflation. Our only assumptions were that $\eta$ and $s$ do not evolve significantly during the non-attractor phase. These assumptions were imposed mainly to keep the analysis under analytical control.  In principle one can go beyond these simplifications and allow the situations in which both $\eta$ and $s$ have significant evolution during the non-attractor phase. Having this said, the setup studied here  in which $ |\eta|, | s| \sim O(1)$ while  $\dot \eta , \dot s \sim 0$  represents large enough class of non-attractor scenarios. 

We have obtained two branches of non-attractor scenarios which can generate  a near scale-invariant power spectrum determined by $\eta = -6 + 5s$ or $\eta =-s$. The first branch encompasses 
the previously known models of non-attractor scenarios in which $s=0$ and $\eta =-6$ \cite{Namjoo:2012aa, Chen:2013aj, Chen:2013eea}.  However, the branch $\eta=-s$ is new which was not noticed previously. This demonstrates the power of EFT of inflation as a very helpful 
setup to  study the paradigm  of inflation model-independently.

We have studied the predictions of the above branches of non-attractor  scenarios for the bispectrum. We have shown that the bispectrum in the new branch $\eta=-s$ has a non-trivial shape.
Depending on the value of the parameter $s$, the bispectrum can change from the equilateral shape 
dominated when $|\cos (2 \pi s)| \rightarrow 1$, to folded shape dominated when $\cos (2 \pi s) \rightarrow 0$. For both configurations, the amplitude of non-Gaussianity is less than unity which is consistent with the observational  constraints. In addition, in this branch $\calR$ is frozen on super-horizon scales and as a result the  non-Gaussianity consistency condition is expected to hold in this branch. We have checked explicitly that this is indeed the case.  On the other hand, for the branch $\eta =-6 + 5 s$, only the local-type non-Gaussianity is generated. The curvature perturbation is not frozen on super-horizon scales and the  non-Gaussianity  consistency condition is violated as previously noticed in particular examples studied 
in \cite{Namjoo:2012aa, Chen:2013aj, Chen:2013eea}. Depending on the value of model parameters
such as $s, c_{s e}, \delta$ and $c_{3 e}$ a large local-type non-Gaussianity is generated. Combined with the Planck's constraints on local-type non-Gaussianity this can be used to constrain the model parameters.  

An interesting conclusion of this model-independent study was that  not all single field non-attractor models can violate the   non-Gaussianity consistency condition. Only in non-attractor models in which  the curvature perturbation is not frozen on super-horizon scales  the  single field  non-Gaussianity consistency condition is  violated. 

\vspace{0.7cm}

{\bf Acknowledgments:}  We would like to thank A. A. Abolhasani, X. Chen, R. Emami, 
E. Komatsu and  M. H. Namjoo   for useful discussions and correspondences.


\appendix
 
\section{The quadratic and cubic actions}
\label{app}

Here we present some details of the quadratic and cubic action. As the non-attractor phase is associated with some paramteres at the  background that evolves fast, one has to be careful in obtaining the actions through expanding these parameters. Note that the mechanism of breaking time-translation invariance is not different from usual attractor phase so we can use result of \cite{Cheung:2007st} prior to expansion which are presented in \eqref{S-matter} and \eqref{EH}.

First, we have to obtain the contributions from $\dot{H}(t+\pi)$ and $H^2(t+\pi)$ to first order of $\epsilon$, since in principle the time derivatives of the Hubble parameter generates powers of $\eta$ which can not be discarded because $|\eta| \sim1$ during the non-attractor phase. These expansions are performed in \eqref{H2} and \eqref{H-dot} . Also other parameters in action may vary fast so we have to expand them and extract their leading behaviors. These terms are $M^4(t+\pi)$ which corresponds to speed of sound and $c_3(t+\pi)$ which is a free parameter of the theory not being fixed by the symmetries. The expansion of these two parameters are presented in \eqref{M4} and \eqref{c3}.

Equipped by expansions of all parameters in the action, the second order action is given as follows
\begin{align}
S_2=\int d^4xa^3 M_p^2 \big \{\epsilon H^2 (\dot{\pi}^2-2N^i\partial_i\pi-2\dot{\pi}\delta N+\delta N^2-\frac{1}{a^2}\partial_i \pi \partial_i \pi )+2\epsilon \eta H^3 \pi \dot{\pi}+\\ \nonumber 
3\epsilon \eta H^4 \pi^2+6\epsilon H^3 \pi \delta N-(1-\frac{1}{c_s^2})\epsilon H^2 [\dot{\pi}^2+\delta N^2 \nonumber 
-2\dot{\pi}\delta N] \\ \nonumber 
+\epsilon \eta^2 H^4 \pi^2-2H\delta N \partial_i N^i+\frac{1}{4}[\partial_iN^j\partial_iN^j-\partial_iN_i \partial_j N_j ]-3H^2\delta N^2 \big \}+{\cal O}(\epsilon^2)
\end{align}
Simplifying the above action yields
\begin{align}
\label{action-app}
S_2=\int d^4xa^3 M_p^2 \epsilon H^2\big \{\frac{\dot{\pi}^2}{c_s^2}+\frac{\delta N^2}{c_s^2}-\frac{2\dot{\pi}\delta N}{c_s^2}+6H\pi \delta N -2N^i\partial_i \pi-\frac{1}{a^2}\partial_i\pi \partial_i \pi \big \}\\ \nonumber 
+\int d^4x a^3 M_p^2 H^2 \big \{ -2\delta N \partial_i N^i -3\delta N^2 \big \}
\end{align}

As usual, $\delta N$ and $N^i$ are non-dynamical in the sense that their equation involves no time derivatives. Solving their equations algebraically we obtain 
\begin{align}
\label{N-Ni}
\delta N=\epsilon H \pi +{\cal O}(\epsilon^2) \nonumber \\ 
\partial_i N^i=-\frac{\epsilon H \dot{\pi}}{c_s^2}+{\cal O}(\epsilon^2)
\end{align}
It is worth mentioning that these results are independent of the attractor or non-attractor assumption. Also the results are independent of the  small or large variation of sound speed as  $s= \dot c_s/H c_s$ does not appear in the quadratic action.  

Having obtained $\delta N$ and $N^i$ one has to plug them back into action (\ref{action-app}) to find the final second order action in terms of $\pi$. As can be seen from Eq. (\ref{N-Ni}) both $\delta N$ and $N^i$ are proportional to $\epsilon$. However, during the non-attractor phase $\epsilon$ falls off exponentially so one expects that the contributions from $\delta N$ and $N^i$ in the action 
(\ref{action-app}) to be sub-leading. Indeed, from (\ref{action-app}) we see that the terms containing 
$\delta N$ and $N^i$ are higher orders in $\epsilon$ compared to terms containing only $\pi$. Therefore, to leading order in power of $\epsilon$, one can safely neglect the contribution from 
$\delta N$ and $N^i$ in the action corresponding to set $\delta N =N^i=0$. This represents our decoupling limit in which we keep only terms of $O(\epsilon)$ and discard terms of higher orders 
in $\epsilon$.  Also in this limit, one can safely neglect the contribution from the gravitational action  \eqref{EH}. In the decoupling action (\ref{action-app}) reduces to Eq. (\ref{action-quad}).

The third order action is complicated in general but in the decoupling limit as described above the action simplifies significantly yielding 
\ba
S_3&=& \int d^4x a^3 \epsilon M_p^2 H^2 \Big [\frac{(\eta-2s)H}{c_s^2}\pi \dot{\pi}^2 -\eta H \frac{1}{a^2}\pi (\partial_i \pi)^2 \nonumber\\
&&~~~~~~~~~~~~~~~~~~~~~~~~~ -(1+\frac{2}{3}c_3)(1-\frac{1}{c_s^2})\dot{\pi}^3+(1-\frac{1}{c_s^2})\dot{\pi}\frac{1}{a^2}(\partial_i \pi)^2 \Big ].
\ea
Note that, unlike second order action, this action is sensitive to large time variation of $c_s$ as $s$ appears directly in the action. In addition the cubic action depends on $c_3$ too.

\bigskip


{}


\begin{thebibliography}{}

\bibitem{Ade:2013lta}
  P.~A.~R.~Ade {\it et al.}  [Planck Collaboration],
  arXiv:1303.5076 [astro-ph.CO].

\bibitem{Ade:2013uln}
  P.~A.~R.~Ade {\it et al.}  [Planck Collaboration],
  arXiv:1303.5082 [astro-ph.CO].
  
  
  
\bibitem{Cheung:2007st} 
  C.~Cheung, P.~Creminelli, A.~L.~Fitzpatrick, J.~Kaplan and L.~Senatore,
  JHEP {\bf 0803}, 014 (2008)
  [arXiv:0709.0293 [hep-th]].


\bibitem{Senatore:2010wk} 
  L.~Senatore and M.~Zaldarriaga,
  JHEP {\bf 1204}, 024 (2012)
  [arXiv:1009.2093 [hep-th]].
  
\bibitem{Ade:2013ydc} 
  P.~A.~R.~Ade {\it et al.}  [Planck Collaboration],
  Astron.\ Astrophys.\  {\bf 571}, A24 (2014)
  [arXiv:1303.5084 [astro-ph.CO]].
  
\bibitem{Alishahiha:2004eh} 
  M.~Alishahiha, E.~Silverstein and D.~Tong,
  Phys.\ Rev.\ D {\bf 70}, 123505 (2004)
  [hep-th/0404084].


\bibitem{Komatsu:2009kd}
  E.~Komatsu {\it et al.},
  ``Non-Gaussianity as a Probe of the Physics of the Primordial Universe and the Astrophysics of the Low Redshift Universe,''
  arXiv:0902.4759 [astro-ph.CO].

\bibitem{Chen:2010xka}
  X.~Chen,
  ``Primordial Non-Gaussianities from Inflation Models,''  Adv.\ Astron.\  {\bf 2010}, 638979 (2010)  [arXiv:1002.1416 [astro-ph.CO]].  

\bibitem{Komatsu:2010hc}
  E.~Komatsu,
  ``Hunting for Primordial Non-Gaussianity in the Cosmic Microwave Background,''
  Class.\ Quant.\ Grav.\  {\bf 27}, 124010 (2010)
  [arXiv:1003.6097 [astro-ph.CO]].


\bibitem{Maldacena:2002vr} 
  J.~M.~Maldacena,
  JHEP {\bf 0305}, 013 (2003)
  [astro-ph/0210603].

\bibitem{Creminelli:2004yq} 
  P.~Creminelli and M.~Zaldarriaga,
  JCAP {\bf 0410}, 006 (2004)
  [astro-ph/0407059].

\bibitem{Pajer:2013ana} 
  E.~Pajer, F.~Schmidt and M.~Zaldarriaga,
  Phys.\ Rev.\ D {\bf 88}, no. 8, 083502 (2013)
  [arXiv:1305.0824 [astro-ph.CO]].

\bibitem{Namjoo:2012aa} 
  M.~H.~Namjoo, H.~Firouzjahi and M.~Sasaki,
  Europhys.\ Lett.\  {\bf 101}, 39001 (2013)
  [arXiv:1210.3692 [astro-ph.CO]].


\bibitem{Chen:2013aj} 
  X.~Chen, H.~Firouzjahi, M.~H.~Namjoo and M.~Sasaki,
  Europhys.\ Lett.\  {\bf 102}, 59001 (2013)
  [arXiv:1301.5699 [hep-th]].

\bibitem{Chen:2013eea} 
  X.~Chen, H.~Firouzjahi, E.~Komatsu, M.~H.~Namjoo and M.~Sasaki,
  JCAP {\bf 1312}, 039 (2013)
  [arXiv:1308.5341 [astro-ph.CO]].

\bibitem{Kinney:2005vj} 
  W.~H.~Kinney,
  Phys.\ Rev.\ D {\bf 72}, 023515 (2005)
  [gr-qc/0503017].

\bibitem{Motohashi:2014ppa} 
  H.~Motohashi, A.~A.~Starobinsky and J.~Yokoyama,
  arXiv:1411.5021 [astro-ph.CO].

\bibitem{ArmendarizPicon:1999rj} 
  C.~Armendariz-Picon, T.~Damour and V.~F.~Mukhanov,
  Phys.\ Lett.\ B {\bf 458}, 209 (1999)
  [hep-th/9904075].

\bibitem{Garriga:1999vw} 
  J.~Garriga and V.~F.~Mukhanov,
  Phys.\ Lett.\ B {\bf 458}, 219 (1999)
  [hep-th/9904176].
  
\bibitem{Huang:2013oya} 
  Q.~G.~Huang and Y.~Wang,
  JCAP {\bf 1306}, 035 (2013)
  [arXiv:1303.4526 [hep-th]].


\bibitem{Cheung:2007sv} 
  C.~Cheung, A.~L.~Fitzpatrick, J.~Kaplan and L.~Senatore,
  JCAP {\bf 0802}, 021 (2008)
  [arXiv:0709.0295 [hep-th]].

\bibitem{Assassi:2012et} 
  V.~Assassi, D.~Baumann and D.~Green,
  JHEP {\bf 1302}, 151 (2013)
  [arXiv:1210.7792 [hep-th]].

\bibitem{Flauger:2013hra} 
  R.~Flauger, D.~Green and R.~A.~Porto,
  JCAP {\bf 1308}, 032 (2013)
  [arXiv:1303.1430 [hep-th]].


\bibitem{Berezhiani:2014tda} 
  L.~Berezhiani, J.~Khoury and J.~Wang,
  JCAP {\bf 1406}, 056 (2014)
  [arXiv:1401.7991 [hep-th]].

\bibitem{Berezhiani:2013ewa} 
  L.~Berezhiani and J.~Khoury,
  JCAP {\bf 1402}, 003 (2014)
  [arXiv:1309.4461 [hep-th]].

\bibitem{Sreenath:2014nca} 
  V.~Sreenath, D.~K.~Hazra and L.~Sriramkumar,
  arXiv:1410.0252 [astro-ph.CO].

\bibitem{Sreenath:2014nka} 
  V.~Sreenath and L.~Sriramkumar,
  JCAP {\bf 1410}, no. 10, 021 (2014)
  [arXiv:1406.1609 [astro-ph.CO]].

\bibitem{Creminelli:2012ed} 
  P.~Creminelli, J.~Norena and M.~Simonovic,
  JCAP {\bf 1207}, 052 (2012)
  [arXiv:1203.4595 [hep-th]].

\bibitem{Creminelli:2013cga} 
  P.~Creminelli, A.~Perko, L.~Senatore, M.~Simonovic and G.~Trevisan,
  JCAP {\bf 1311}, 015 (2013)
  [arXiv:1307.0503 [astro-ph.CO]].

\bibitem{Dimastrogiovanni:2014ina} 
  E.~Dimastrogiovanni, M.~Fasiello, D.~Jeong and M.~Kamionkowski,
  arXiv:1407.8204 [astro-ph.CO].













\bibitem{Chen:2009zp} 
  X.~Chen and Y.~Wang,
  JCAP {\bf 1004}, 027 (2010)
  [arXiv:0911.3380 [hep-th]].


\bibitem{Chen:2009we} 
  X.~Chen and Y.~Wang,
  Phys.\ Rev.\ D {\bf 81}, 063511 (2010)
  [arXiv:0909.0496 [astro-ph.CO]].

\bibitem{Emami:2013lma} 
  R.~Emami,
  JCAP {\bf 1404}, 031 (2014)
  [arXiv:1311.0184 [hep-th]].

\bibitem{Pi:2012gf} 
  S.~Pi and M.~Sasaki,
  JCAP {\bf 1210}, 051 (2012)
  [arXiv:1205.0161 [hep-th]].





\bibitem{Chen:2006nt} 
  X.~Chen, M.~x.~Huang, S.~Kachru and G.~Shiu,
  JCAP {\bf 0701}, 002 (2007)
  [hep-th/0605045].

\bibitem{Shandera:2006ax} 
  S.~E.~Shandera and S.-H.~H.~Tye,
  JCAP {\bf 0605}, 007 (2006)
  [hep-th/0601099].


\bibitem{Weinberg:2005vy}
  S.~Weinberg,
  Phys.\ Rev.\ D {\bf 72}, 043514 (2005)
  [hep-th/0506236].

 
\bibitem{Wang:2013zva}
  Y.~Wang,
  arXiv:1303.1523 [hep-th].
  

\end{thebibliography}
\end{document}